\documentstyle[doublespacing]{mn}
\begin{document}
{\title [Primeval Galaxies in the
HDF]{Candidate Primeval Galaxies in the Hubble Deep Field}}

\author[D.L. Clements \& W.J. Couch]{D.L. Clements,$^1$ and W.J.
Couch$^{1,2}$\\
$^1$European Southern Observatory, Karl-Schwarzschild-Strasse 2,
D-85748 Garching-bei-Munchen, Germany\\
$^2$School of Physics, University of New South Wales, Sydney, NSW
2052, Australia
}
\maketitle

\begin{abstract}
We present the results of colour-selection of candidate high redshift
(2.6 $<$ z $<$ 3.9) galaxies within the Hubble Deep Field based on the
Ly-break at 912\AA . We find 8 such objects in the region, giving a
comoving number density comparable to that of nearby bright
galaxies (for a flat q$_0$=0.5, H$_0$=100 kms$^{-1}$Mpc$^{-1}$
universe). We provide basic data on the properties of these objects,
and show that despite their absolute magnitude being significantly
brighter than L$^*$ (typically M$_B$ = -22), they are generally smaller
than nearby galaxies. Furthermore, visual inspection of their images
shows that they are all highly disturbed systems, with multiple
nuclei, tails and plumes, suggesting that they are undergoing merging
processes similar to most nearby starburst galaxies.  Theoretical
models suggest that galaxies form by accumulation of numerous
subcomponents, and we suggest that we are seeing this process underway
in these objects. It is thus possible that the epoch of galaxy formation
might have been discovered.
\end{abstract}

\begin{keywords}
galaxies:formation - galaxies:primeval - galaxies:interactions -
galaxies:starburst
\end{keywords}

\section{Introduction}

The discovery of very high redshift and primeval galaxies has been a
`holy-grail' for cosmologists for many years \cite{p74}. Amongst
numerous approaches to finding forming galaxies and understanding
their subsequent evolution, faint imaging surveys, for both number
counts \cite{t88} and to find individual interesting objects selected
by various colour criteria (e.g. multicolour searches for high
redshift quasars) are well established techniques \cite{s94}. Among
the more interesting results to emerge from these studies has been the
discovery, at blue wavelengths, of many more galaxies than standard
`no--evolution' models predict -- the so--called `faint blue excess'
\cite{t88}. However, various spectroscopic (Glazebrook et al. \shortcite{g95a},
Lilly et al. \shortcite{l95}) and HST-based morphological studies
(Driver et al. \shortcite{d95}, Glazebrook et al. \shortcite{g95b})
have shown this excess population to reside at only intermediate redshifts 
($z<1$), and to consist of dwarf late-type
spiral/irregular galaxies which undergo substantial luminosity and
colour evolution between $z\sim 1$ and the present time. On the other hand, 
these same studies have shown the redder early-type (E/S0abc) population 
to have undergone little if any evolution in
this period.  The formation epoch of these galaxies, therefore, must lie
at considerably greater redshift, even if the nature of the faint blue
excess has now been determined.

In this context, the Director of the Hubble Space Telescope (HST)
assigned a substantial allocation of discretionary time ($\sim$ 150
orbits) to a project aimed at obtaining very deep Wide-Field \&
Planetary Camera 2 (WFPC-2) images of a single `random' field
(position 12h 36m 49.4s +62d 12' 58'', J2000 \cite{hdf1}; hereafter
referred to as the `Hubble Deep Field' (HDF)). Four filters were
chosen for these observations, F300W, F450W, F606W and F814W. The
data, which were taken over the 13-day period 1995 December 18--30,
were made public immediately \cite{w96} and we present here the
results of a programme to search for high redshift counterparts of
normal galaxies using the presence of the Lyman break at 912\AA ~in
the F300W filter as an indicator of redshift.

In the next section we describe our selection technique. In section 3
we present our results, while in section 4 we discuss the properties
of the objects we find.  We draw our conclusions in section 5.

\section{Selection of High Redshift Objects}

The HDF extends to magnitude limits (on the STMAG system\footnote{
STMAG~=~-2.5log$_{10}f_{\lambda}$~-~21.10 where $f_{\lambda}$ is the
flux in ergs\,cm$^{-1}$\,s$^{-1}$\,\AA$^{-1}$ received in the
bandpass}) of 26.7, 28.8, 30.1, and 30.3 in the F300W, F450W, F606W,
and F814W filters, respectively, for a 5 $\sigma$ detection in 16
connected 0.04\arcsec ~pixels \cite{hdf2}.  For the present study we
wish to detect and then examine the morphological properties of high
redshift galaxies. We identify high redshift candidates by looking for
a substantial magnitude difference between the F300W and F450W
filters.  For galaxies with 2.6 $<$z$<$3.9, the F300W passband will
correspond to rest-frame wavelengths shorter than the Lyman cutoff at
912\AA ~and the F450W filter to longer wavelengths. These high
redshift galaxies will thus be undetectable at F300W or will have
anomalously red F300W - F450W colours. Objects at still higher
redshift will additionally have F450W suppressed.

A similar method has been used successfully by Steidel et al
\shortcite{s92} to look for companions to high redshift quasars and
damped absorption-line systems, and by Guhathakurta et al
\shortcite{g90} to set a limit on the redshift of the
faint-blue-galaxy population. Additionally, the spectral energy
distribution of objects at high redshift is likely to be fairly flat
redward of the Lyman cutoff \cite{d92},  an assumption that is
borne out by recent Keck spectroscopy of z$\sim$ 3 galaxies
\cite{s96}. Our selection technique is thus to look for all those
objects with F300W - F450W $>$ 2, and remaining colours (F450W - F606W
and F606W - F814W) between -0.6 and 0.6. Consideration of the
colour-colour diagram in \cite{st95} and the detailed photometric
modeling of Fukugita et al \shortcite{f95} demonstrate that it is
very unlikely for lower redshift galaxies of any type to have such
colours. Objects not detected in the F300W filter are treated as if
they have a magnitude of 27.7, corresponding to the 2$\sigma$ upper
limit, for the purposes of this selection. We are thus constrained to
objects with an F450W magnitude of 25.7 or brighter for our
candidates, which usefully guarantees that they will be bright enough
for us to study their morphology.  Examining the flux zero points for
these filters shows that for an F300W - F450W colour $>$ 2, the flux
ratio between the F450W and F300W bands is $>$ 6. This is higher than
could be caused by the 4000 \AA break or by absorption in the
Ly$\alpha$ forest where the continuum suppression factor at these
redshifts is only about 50\% \cite{sl94}.

This selection was based on object catalogs derived from the
combined `drizzled' HDF images \cite{w96} using the automated
photometry program SeXtractor written by E. Bertin. A detailed
description of the techniques this program employs in the detection and
photometry of objects is given by \cite{b96}; its specific application to 
HST WFPC-2 images is discussed by \cite{sm96}. In 
using it here, SeXtractor was first run on a combined F606W+F814W image
to provide a deep `master' list of objects. It was then
run on each of the individual F300W, F450W, F606W, and F814W images, 
the resulting catalogs being matched and merged with the master version.
In all cases, a detection threshold of 1.3$\sigma$ (where $\sigma$ is
the standard deviation of the background noise distribution) and a
minimum area after convolution with a 0.3\,arcsec top--hat filter of
0.05\,arcsec$^{2}$ (30 connected pixels) were adopted. This resulted in an 
average of 200, 420, 510, and 420 objects being detected on each WFC chip in 
F300W, F450W, F606W, and F814W, respectively.

SeXtractor computes several different types of magnitude for each
detected object but we worked solely with the corrected isophotal
(MAG\_ISOCOR) values as they appeared to be the most stable at the
faintest limits.  This particular magnitude provides an estimate of
the total flux from an object by taking the light measured within the
threshold isophote and then, by assuming a gaussian profile, making a
small ($<5$\%) correction for the light lost outside this isophote. As
a check, these magnitudes were compared with the `total' magnitudes
available in catalogs generated at the Space Telescope Science
Institute using the FOCAS package \cite{v93}. With the exception of
the F300W band (see below), excellent agreement was found between the
two, both in zero point and scale.  The scatter observed between the
two sets of magnitudes was also consistent with the random photometric
errors expected from photon statistics.

It was noticed that SeXtractor and FOCAS both had difficulties in
reliably detecting and measuring objects on the F300W images. A visual
inspection revealed that many of the detected objects were spurious
and yet a considerable number of real objects were missed. This is
undoubtedly due to the problems experienced in flat-fielding the F300W
data and removing scattered light.  A conspicuous checker-board
pattern peculiar to the drizzling scheme was also present around the
edges of the frames. These problems were dealt with by using the IRAF
APPHOT routine to measure aperture magnitudes at each of the positions
in our master catalogue, setting the aperture diameter equal to the
major axis length computed for each image on the combined F606W+F814W
frame. It was these F300W aperture magnitudes that were used in the
selection process.

\section{Results}

Using the above colour and magnitude criteria, we searched our
SExtractor catalogue for candidate high redshift objects.
We found 9 candidates at F450W$<$25.7 which satisfied our colour criteria. 
Visual inspection of all these candidates led us to reject one of these
because it lay near the edge of one of the CCD chips and was severely
contaminated by spurious flux at the edge of the dithering pattern.

Basic data on the objects is presented in Table 1. Images of the
objects are shown in Figure 1a, using the combined F606W+F814W frame.
Corresponding F300W images are shown in Figure 1b, demonstrating the
lack of any detections, and thus the presumed presence of the Ly break
in this band.

We also used a similar search technique to find candidate objects at
still higher redshift by looking for a flat F606W -F814W colour
combined with F450W-F606W $>2$ and little or no flux in the F300W
band. This would select for objects with 3.9$<$z$<$5, though with
additional complications from the Ly$\alpha$ forest absorption which
becomes stronger. No such objects were found. 

\begin{figure*}
\vspace{220mm}
\label{images}
\caption{Images of the Objects. The objects are shown in order
from left to right then down the page, starting with 2P1 in the top left and
ending with 4P2 in the bottom middle. All images are 4\arcsec  square.
{\em a} Coadded F606W + F814W image. {\em b} F300W image showing lack
of flux in this band for all objects.}
\end{figure*}

\begin{figure}
\vspace{5.5cm}
\label{mags}
\caption{Histogram of Absolute Magnitudes. These become 1.4 magnitudes
brighter for a q$_0$ = 0.05 universe.}
\end{figure}

\begin{figure}
\vspace{5.5cm}
\caption{Histogram of Object Sizes. These increase by a factor of 2
for a q$_0$ = 0.05 universe.}
\label{sizes}
\end{figure}

\section{The Properties of Candidate z$\sim$3 Galaxies}

We found 8 objects on the 3 WFC chips likely to be in the range 
2.6$<$z$<$3.9. We now calculate the number-density, absolute magnitude
and physical size of these objects; in doing so, we assume them all to
be at the mid-point of this redshift range (ie. $z=3.25$) and adopt
an H$_0$=100, q$_0$=0.5 (i.e. Einstein-de-Sitter) cosmology except
where stated otherwise.

A total of 8 objects within the 4.7 sq. arcmin area of the HDF gives
a surface density for these objects of 1.7 arcmin$^{-2}$. At matching
magnitudes we find a similar surface density of 0.4
arcmin$^{-2}$ to that of the R$\sim$ 25 3$<$z$<3.5$ galaxies
discussed by Steidel et al \cite{s96}.  For the assumed redshift and
cosmology, the number density of high redshift galaxies in the HDF
corresponds to a comoving number density of 0.005 h$^3$Mpc$^{-3}$
(where h$=$ H$_0$/100); in
comparison, nearby bright ($L > L^{*}$) galaxies have a number density
of 0.015 Mpc$^{-3}$ \cite{p94}. The similarity of these two numbers
suggests that we may finally be seeing the progenitors of nearby
bright galaxies. The corresponding number for a q$_0$=0.05 universe is
0.0009 h$^3$Mpc$^{-3}$.

Absolute `blue' magnitudes were determined from the
apparent F450W magnitudes using the K-correction calculated by Cowie
et al. \shortcite{c94} for a star-forming (`spiral') galaxy at this
redshift. This is based on local galaxy SEDs from Coleman, Wu \&
Weedman (1980) in the optical, and Mobasher, Ellis \& Sharples
\shortcite{m86} in the infrared. We also apply the +0.4 magnitude
correction between the F450W and standard B band appropriate for a
flat spectrum object using the information provided by Holtzman et al
\shortcite{h95}. These absolute magnitudes must be treated with great
caution as neither the applicability of this particular K-correction
to our sample of high-redshift candidates nor the required knowledge
of the far-UV spectral energy distribution of different present-day
types are well determined. Indeed the K correction is similarly large
and uncertain for all the filters discussed here since the objects lie
at such a high redshift. This notwithstanding, it is of much interest
that the values found are at the upper end of the luminosity function
of local galaxies\footnote{We note that this calculation is very
sensitive to q$_0$ with our values being about 1.4 magnitudes brighter
if a value of q$_0$=0.05 is adopted.} (see Figure 2). In Figure 3 we
show the sizes of the major axis of the objects measured at an
isophote 3 magnitudes fainter than the peak surface brightness. In
contrast to the luminosities, these appear to be significantly lower
(median diameter $\sim$4\,kpc) than those of their present-day
counterparts. A similarly small-sized population of sources was noted
by Dressler et al. \shortcite{d94} in a galaxy cluster found
serendipitously on a WFPC-2 image and thought to be associated with a
QSO at z=2.2.

\begin{table*}
\label{tab1}
\begin{minipage}{180mm}
\begin{tabular}{ccccccccc}
Name&RA(J2000)&Dec(J2000)&F450W&F606W&F814W&Abs. B
Mag&Size (kpc)\\ \hline
2P1&12:36:45.29&62:13:47.9&25.6&25.6&26.1&-21.6&4.5\\
2P2&12:36:53.34&62:13:30.4&25.3&25.1&25.4&-21.9&4.2\\
2P3&12:36:51.97&62:14:34.8&25.6&26.1&26.6&-21.6&4.0\\
2P4&12:36:55.07&62:13:49.1&25.4&25.7&26.2&-21.8&3.0\\
3P1&12:36:49.85&62:12:44.6&25.4&25.9&26.5&-21.8&5.0\\
3P2&12:36:58.20&62:12:17.7&25.5&25.8&26.2&-21.7&3.9\\
4P1&12:36:45.25&62:11:54.5&23.9&23.8&24.2&-23.4&10.9\\
4P2&12:36:41.58&62:12:39.9&24.5&25.0&25.4&-22.7&3.1\\
\end{tabular}
\caption{Candidate Primeval Galaxies in the HDF.}
All magnitudes are on
the ST magnitude system. No galaxies were detected in the F300W
filter, giving a 2 $\sigma$ magnitude limit $> \sim$27.7. Names follow
the pattern chip no. P object no. so that the first object detected on
chip 2 is 2P1. Absolute magnitudes are calculated assuming a K
correction appropriate for spiral galaxies and the appropriate
magnitude shift from F450W to B given in \cite{h95}.
Values are given for H$_0$ = 100 km$^{-1}$Mpc${-1}$ and q$_0$=0.5. For
q$_0$ = 0.05, absolute magnitudes get brighter by 1.4 mags, and sizes
increase by a factor of 2, while for H$_0$=50, corresponding figures
are 1.5 mags brighter and 2 times bigger.
\end{minipage}
\end{table*}

The object selection is sufficiently bright in the longer wavelength
filters that the images are deep enough to allow examination of the
morphologies of the selected galaxies. Inspection of the images in
Figure 1a immediately demonstrates that all of these objects are
disturbed systems, with multiple nuclei, jets, plumes, and asymmetric
flux distributions. Indeed, the structures observed are similar to
interacting/merging galaxies observed locally e.g. Clements et
al. \shortcite{c96}. It has been known for some time that interactions
and mergers between galaxies can trigger bursts of star formation
\cite{j85}. These morphological signatures are also common at
intermediate redshifts where HST and other high-resolution
observations have revealed objects of this type amongst the blue
`Butcher-Oemler' population in clusters (Couch et
al. \shortcite{co94}, Dressler et al. \shortcite{d94}), the faint blue
field population \cite{g95b}, and mJy radio sources \cite{w95}.

Theoretical models of galaxy formation and evolution (e.g. Kauffman
\shortcite{k95}) currently favour the idea that the galaxies we see
today were formed as a result of the merging/accretion of a number of
subunits. If this is the case then each such accretion event is likely
to be accompanied by the vigorous star formation we see in local
mergers. The disturbed, multiple-component nature of the HDF high
redshift galaxies presented here suggests that this is a wide-spread
occurrence at these very early epochs. The close match between the
number densities of these objects and local galaxies might indicate
that we are at last seeing the epoch where most of the galaxies we see
today were being assembled from such subunits. Alternatively, if there
has been considerable number-density evolution in galaxies at lower
redshifts (0.5 -- 1), as might be indicated by the faint-blue-galaxy
population, then we might just be seeing occasional starbursts in a
population of more numerous small galaxies.

The effect of viewing these objects in their restframe UV ($\sim$ 1670
\AA) is also another factor that must be considered. We might just be
seeing the hottest, brightest regions of active star formation in a
larger, lower surface brightness structure that we cannot see.
Observations of the HDF at IR wavelengths, particularly those
conducted with the new NICMOS camera soon to be installed on HST, will
be invaluable in addressing the problem of the observed restframe
wavelength. For example, the centre of the $K$ window corresponds, at
these redshifts, to $\sim$ 5100 \AA , conveniently in the restframe
optical emission. Observations at these wavelengths from the ground
will also be useful in better constraining the spectral nature of
these objects; we shall be conducting these in the forthcoming months.

\section{Conclusions}
We have presented results on a sample of objects very likely to lie at
2.6$<$z$<$3.9. We find that these objects have similar co-moving
number density to nearby bright galaxies. However they are
significantly more luminous and smaller than nearby galaxies. Almost
all of our high redshift objects show signs of a disturbed and
multi--component morphology, which we interpret as being due to
merging processes taking place during the assembly of a galaxy.
\\ ~\\
{\bf ACKNOWLEDGMENTS}
\\ ~\\
We congratulate the team that devised and conducted the HDF
observations on a job well done, and are especially grateful to the
Director of the HST, Bob Williams, for organizing such an excellent
programme. We would also like to thank the ESO/ECF HDF group for many
useful discussions, and especially Bob Fosbury, Richard Hook, Rene
Mendez and Hans-Martin Adorf for their help. We also thank the
anonymous referee for many helpful comments. DLC is funded by an ESO
Fellowship and WJC by an ESO Visitors position.

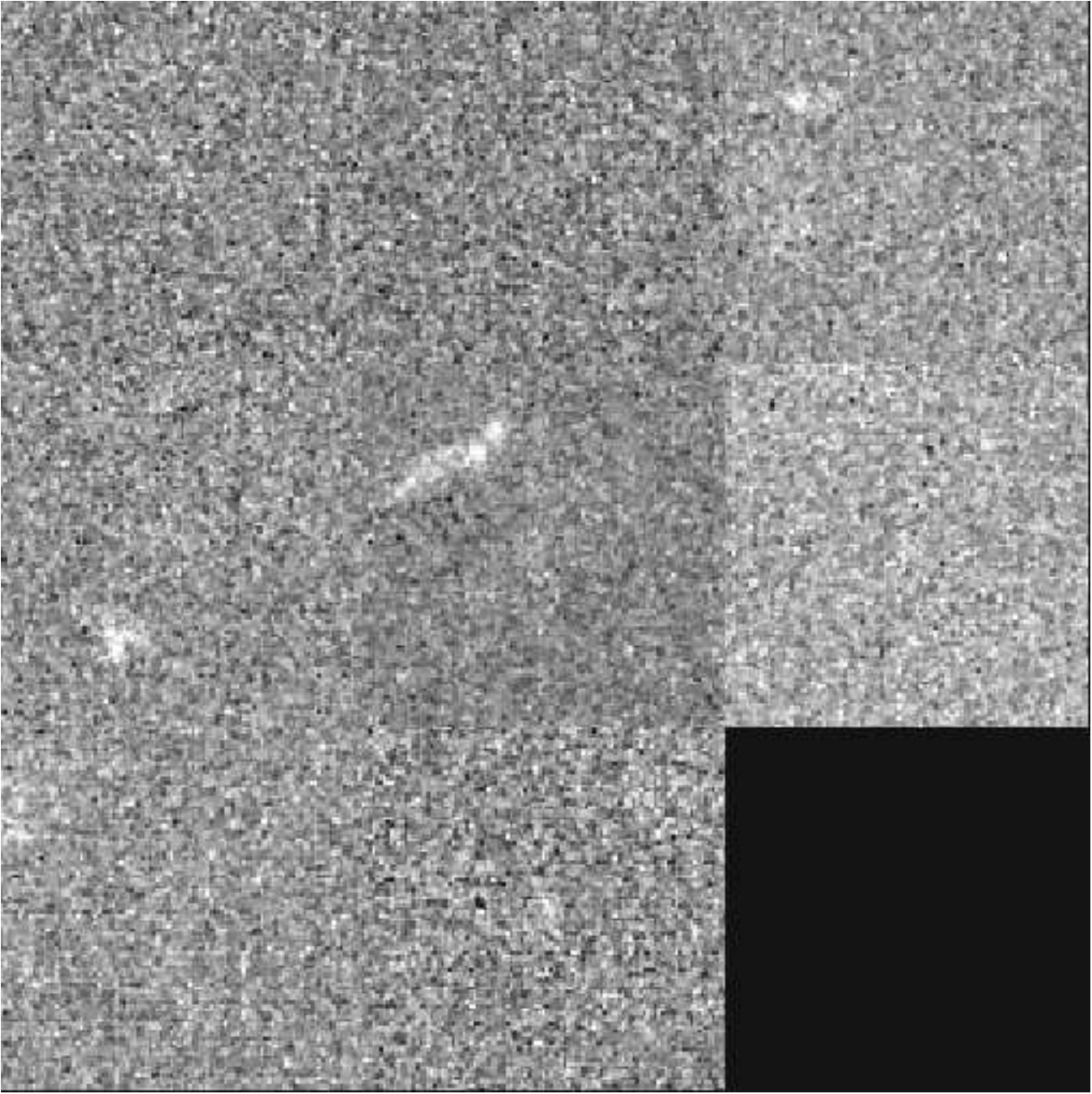
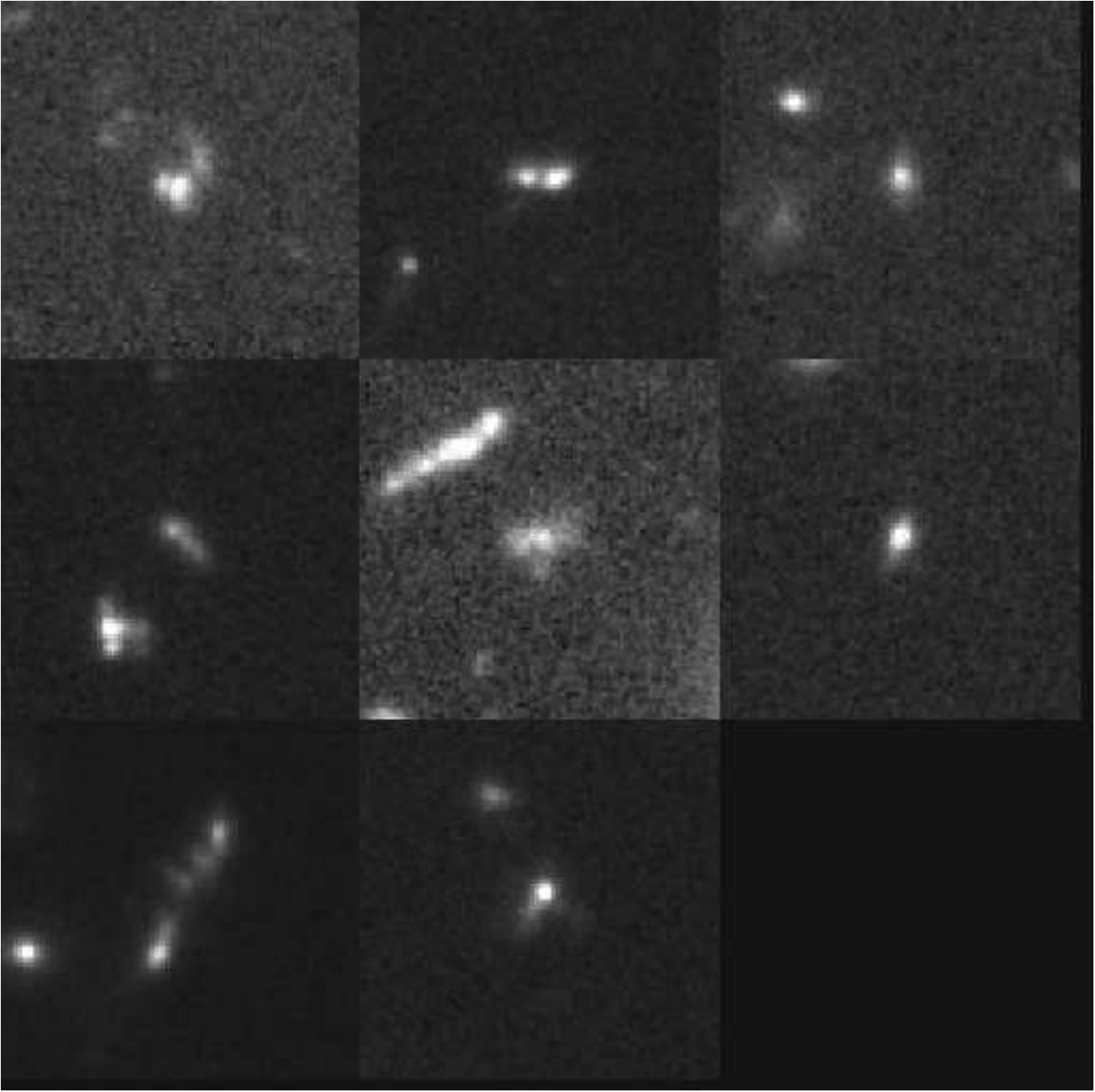
\end{document}